\documentclass[%
reprint,
superscriptaddress,longbibliography
amsmath,amssymb,
aps,
]{revtex4-2}

\usepackage{graphicx}
\usepackage{dcolumn}
\usepackage{bm}
\usepackage{graphicx}
\usepackage{color,soul}
\usepackage{amsmath,amssymb}
\usepackage{mathrsfs}
\usepackage{color}
\usepackage{booktabs}
\usepackage{afterpage}
\usepackage{pifont}
\usepackage{xcolor}
\usepackage[version=3]{mhchem}
\usepackage{natbib}
\usepackage{soul}
\usepackage[caption=false]{subfig}
\usepackage{xtab,afterpage}
\usepackage{lipsum}
\usepackage{array}
\usepackage{multirow}

\begin{document}
	
	\preprint{APS/123-QED}
	
	\title{A Universal Crystal-Field Design Principle for Orbital-Order-Driven Altermagnetism}
	
	\author{Shantanu Pathak}
	\email{shantanu.pathak@physics.iitd.ac.in}
	
	\author{Saswata Bhattacharya}
	\email{saswata@physics.iitd.ac.in}
	
	\affiliation{Department of Physics, Indian Institute of Technology Delhi,
		New Delhi 110016, India}
	
	\date{\today}

\begin{abstract}
	Altermagnets combine collinear antiferromagnetic order with nonrelativistic spin splitting, enabling spintronic functionalities without relying on spin--orbit coupling. While staggered orbital ordering has recently emerged as an alternative route to altermagnetism, its generality has remained unexplored. Here, we establish a universal crystal-field design principle for orbital-order-driven altermagnetism. We show that structural relaxation consistently reconstructs the crystal-field landscape, activating a common $d_{xz}/d_{yz}$ orbital manifold that drives spontaneous staggered orbital ordering and robust $d$-wave nonrelativistic spin splitting across transition-metal compounds spanning electron fillings from $d^1$ to $d^7$. By introducing a unified symmetry framework based on layer-dependent magnetic and orbital order parameters, we demonstrate how interlayer stacking determines whether the system realizes a bulk altermagnetic state or a globally compensated antialtermagnetic phase. Furthermore, we reveal that this symmetry-protected spin-split texture gives rise to highly anisotropic spin-polarized conductivities. Our results establish crystal-field engineering as a predictive design strategy for discovering and engineering orbital-order-driven altermagnets.
\end{abstract}
	
	
	\maketitle
	
\section*{Introduction}
The recent discovery of altermagnetism has established a fundamentally distinct magnetic phase that bridges the traditional dichotomy between ferromagnetism and antiferromagnetism \cite{vsmejkal2022beyond, jungwirth2024altermagnets, gonzalez2021efficient, bailing}. Altermagnets combine the vanishing net magnetization of collinear antiferromagnets with a nonrelativistic spin splitting (NRSS) of electronic bands, thereby enabling highly spin-polarized transport without relying on relativistic spin--orbit coupling \cite{vsmejkal2022beyond, Denisov2024PRB_NRSDynamics, Fang2024PRL_QGAM, w52v-blqm, ss_prb}. Owing to this unique combination of ultrafast magnetic dynamics, robustness against external magnetic fields, and efficient spin-current generation, altermagnets have rapidly emerged as a promising platform for next-generation spintronic technologies \cite{sinova2015spin, vzutic2004spintronics, Pan2024PRL_GST, Samanta2025NanoLett_filter, feng2022anomalous}.

The microscopic origin of altermagnetism has traditionally been understood through the interplay between rigid crystal symmetry and magnetic order, where specific crystallographic point groups dictate the characteristic momentum-dependent spin splitting \cite{vsmejkal2022beyond, pathak2026strain}. More recently, staggered orbital ordering (OO) has been recognized as an alternative microscopic route to altermagnetism by breaking the translational equivalence between magnetic sublattices \cite{quintin, PhysRevLett.132.236701, Zhang2026NanoLett_hidden}. Although this important advance broadens the landscape of candidate materials beyond conventional symmetry-protected altermagnets, two fundamental questions remain unanswered. Is orbital-order-driven altermagnetism restricted to a few isolated material families, or does it represent a general microscopic mechanism? More importantly, can one formulate predictive design rules that link crystal-field reconstruction, orbital ordering, magnetic symmetry, and spin transport across chemically diverse compounds? To date, however, no unified microscopic framework has connected these ingredients into a transferable design strategy.

In this work, we answer these questions by establishing a universal crystal-field design principle for orbital-order-driven altermagnetism. We demonstrate that structural relaxation naturally reconstructs the crystal-field landscape, stabilizing a common active $d_{xz}/d_{yz}$ orbital manifold that spontaneously develops staggered orbital ordering. This crystal-field reconstruction, rather than the specific chemical composition or lattice geometry, provides the fundamental microscopic origin of the nonrelativistic spin splitting. By introducing layer-dependent magnetic and orbital order parameters, $L_i$ and $\Lambda_i$, we formulate a unified symmetry framework that determines whether the layer-resolved spin splitting survives as a bulk altermagnetic state or becomes compensated into an antialtermagnetic phase through interlayer stacking.

To demonstrate the generality of this framework, we systematically investigate transition-metal compounds spanning diverse electron fillings, magnetic configurations, and crystal structures. Representative systems are discussed in the main text, while additional compounds and the complete Ruddlesden–Popper (RP) homologous series are presented in the Supplemental Material \cite{SM}, demonstrating the robustness of the proposed mechanism across different dimensionalities and crystal structures.

Finally, employing semiclassical Boltzmann transport theory \cite{callaway2013quantum}, we show that this universal symmetry mechanism directly gives rise to highly anisotropic spin-polarized conductivities. Our results elevate orbital ordering from a material-specific phenomenon to a predictive design principle for altermagnetism. Rather than viewing orbital-order-driven altermagnets as isolated material realizations, we establish a unified framework linking crystal-field reconstruction, orbital ordering, symmetry, and spin transport, thereby providing predictive design rules for discovering and engineering altermagnetic materials across a broad range of transition-metal compounds.

\section*{Computational Methods}
First-principles calculations were carried out within the framework of density functional theory (DFT) using the Vienna \textit{Ab initio} Simulation Package (VASP) \cite{kresse1996vasp1, kresse1996vasp2}. The interactions between valence electrons and ionic cores were described using the projector augmented-wave (PAW) method \cite{blochl1994projector}, while exchange--correlation effects were treated within the generalized-gradient approximation (GGA) using the Perdew--Burke--Ernzerhof (PBE) functional \cite{perdew1996generalized}. Strong electronic correlations associated with the transition-metal $d$ orbitals were incorporated using the rotationally invariant GGA+$U$ formalism \cite{dudarev1998electron}.

All crystal structures were fully optimized until the total energies and atomic forces satisfied stringent convergence criteria. Material-specific Hubbard $U$ values were adopted from previous experimental and theoretical studies to reproduce the appropriate electronic and magnetic ground states. Details of the convergence parameters, $U$ values, van der Waals corrections, and structural optimization procedures are summarized in the Supplemental Material.

Electronic transport properties were calculated by constructing maximally localized Wannier functions (MLWFs) using the \textsc{Wannier90} package \cite{mostofi2008wannier90}. The resulting tight-binding Hamiltonians were subsequently employed within the \textsc{WannierBerri} \cite{tsirkin2021high} framework to evaluate nonrelativistic charge and spin conductivities using the Kubo formalism. Magnetic spin-group symmetries together with tensor transformation rules were analyzed using the \textsc{FINDSYM} \cite{stokes2005findsym}, \textsc{FindSpinGroup} \cite{chen2024_spingroup_prb110, chen2024_spinsg_prx}, Bilbao Crystallographic Server~\cite{aroyo2006bilbao,aroyo2011crystallography}, and \textsc{TensorSymmetry} \cite{xiao2025tensorsymmetry} packages.

\begin{table*}[t]
	\caption{
		Summary of the proposed crystal-field design principle for orbital-order-driven altermagnetism. Despite spanning diverse transition-metal compounds and electron fillings, all investigated systems follow the same microscopic pathway: crystal-field reconstruction induces staggered orbital ordering, generating layer-resolved nonrelativistic spin splitting. The interlayer symmetry subsequently determines whether the resulting bulk state is altermagnetic (AM), antialtermagnetic (anti-AM) or ferri-altermagnetic (ferri-AM). Within the conventional crystal-field picture, LS (HS) states are favored when the octahedral crystal-field splitting $\Delta_{\mathrm O}$ is larger (smaller) than the electron-pairing energy $P$.
	}
	\label{tab:materials}
	
	\centering
	\renewcommand{\arraystretch}{1.15}
	\begin{tabular}{lccccc}
		\hline\hline
		Material &
		Electron
		filling &
		Crystal-field-selected active manifold &
		Orbital
		ordering &
		Interlayer
		symmetry &
		Bulk state$^*$
		\\
		\hline
		
		SrVO$_3$
		&
		$d^{1}$
		&
		$d_{xz}/d_{yz}$
		&
		$\Lambda_i\neq0$
		&
		$t^{\prime}\mathcal{RT}$/$t^{\prime}\mathcal{T}$
		&
		AM/anti-AM
		\\
		
		Sr$_2$VO$_4$
		&
		$d^{1}$
		&
		$d_{xz}/d_{yz}$
		&
		$\Lambda_i\neq0$
		&
		$t^{\prime}\mathcal{RT}$/$t^{\prime}\mathcal{T}$
		&
		AM/anti-AM
		\\
		
		Sr$_3$V$_2$O$_7$
		&
		$d^{1}$
		&
		$d_{xz}/d_{yz}$
		&
		$\Lambda_i\neq0$
		&
		$t^{\prime}\mathcal{T}$
		&
		anti-AM
		\\
		
		Sr$_4$V$_3$O$_{10}$
		&
		$d^{1}$
		&
		$d_{xz}/d_{yz}$
		&
		$\Lambda_i\neq0$
		&
		$t^{\prime}\mathcal{RT}$
		&
		ferri-AM
		\\
		
		LaVO$_3$
		&
		$d^{2}$
		&
		$d_{xz}/d_{yz}$
		&
		$\Lambda_i\neq0$
		&
		$t^{\prime}\mathcal{RT}$/$t^{\prime}\mathcal{T}$
		&
		AM/anti-AM
		\\
		
		SrMoO$_3$
		&
		$d^{2}$
		&
		$d_{xz}/d_{yz}$
		&
		$\Lambda_i\neq0$
		&
		$t^{\prime}\mathcal{RT}$/$t^{\prime}\mathcal{T}$
		&
		AM/anti-AM
		\\
		
		Sr$_2$MoO$_4$
		&
		$d^{2}$
		&
		$d_{xz}/d_{yz}$
		&
		$\Lambda_i\neq0$
		&
		$t^{\prime}\mathcal{RT}$/$t^{\prime}\mathcal{T}$
		&
		AM/anti-AM
		\\
		
		Sr$_3$Mo$_2$O$_7$
		&
		$d^{2}$
		&
		$d_{xz}/d_{yz}$
		&
		$\Lambda_i\neq0$
		&
		$t^{\prime}\mathcal{T}$
		&
		anti-AM
		\\
		
		Sr$_4$Mo$_3$O$_{10}$
		&
		$d^{2}$
		&
		$d_{xz}/d_{yz}$
		&
		$\Lambda_i\neq0$
		&
		$t^{\prime}\mathcal{RT}$
		&
		ferri-AM
		\\

		SrRuO$_3$
		&
		LS-$d^{4}$
		&
		$d_{xz}/d_{yz}$
		&
		$\Lambda_i\neq0$
		&
		$t^{\prime}\mathcal{RT}$/$t^{\prime}\mathcal{T}$
		&
		AM/anti-AM
		\\
		
		Sr$_2$RuO$_4$
		&
		LS-$d^{4}$
		&
		$d_{xz}/d_{yz}$
		&
		$\Lambda_i\neq0$
		&
		$t^{\prime}\mathcal{RT}$/$t^{\prime}\mathcal{T}$
		&
		AM/anti-AM
		\\
		
		Sr$_3$Ru$_2$O$_7$
		&
		LS-$d^{4}$
		&
		$d_{xz}/d_{yz}$
		&
		$\Lambda_i\neq0$
		&
		$t^{\prime}\mathcal{T}$
		&
		anti-AM
		\\
		
		Sr$_4$Ru$_3$O$_{10}$
		&
		LS-$d^{4}$
		&
		$d_{xz}/d_{yz}$
		&
		$\Lambda_i\neq0$
		&
		$t^{\prime}\mathcal{RT}$
		&
		ferri-AM
		\\
		
		Sr$_2$RhO$_4$
		&
		LS-$d^{5}$
		&
		$d_{xz}/d_{yz}$
		&
		$\Lambda_i\neq0$
		&
		$t^{\prime}\mathcal{RT}$/$t^{\prime}\mathcal{T}$
		&
		AM/anti-AM
		\\
		
		Sr$_2$IrO$_4$
		&
		LS-$d^{5}$
		&
		$d_{xz}/d_{yz}$
		&
		$\Lambda_i\neq0$
		&
		$t^{\prime}\mathcal{RT}$
		&
		AM
		\\
		
		KFeF$_3$
		&
		HS-$d^{6}$
		&
		$d_{xz}/d_{yz}$
		&
		$\Lambda_i\neq0$
		&
		$t^{\prime}\mathcal{RT}$
		&
		AM
		\\
		
		KCoF$_3$
		&
		HS-$d^{7}$
		&
		$d_{xz}/d_{yz}$
		&
		$\Lambda_i\neq0$
		&
		$t^{\prime}\mathcal{RT}$
		&
		AM
		\\
		
		K$_2$CoF$_4$
		&
		HS-$d^{7}$
		&
		$d_{xz}/d_{yz}$
		&
		$\Lambda_i\neq0$
		&
		$t^{\prime}\mathcal{RT}$
		&
		AM
		\\
		
		\hline\hline
	\end{tabular}
	\begin{flushleft}
		*The relative energies of all orbital-ordering configurations are summarized in Supplementary Table~S1.
	\end{flushleft}
\end{table*}
\section*{Results and Discussions}
\subsection*{Universal symmetry framework for orbital-order-driven altermagnetism}
The central question addressed in this work is not whether orbital ordering can induce altermagnetism in a particular material, but rather whether a common microscopic principle governs orbital-order-driven nonrelativistic spin splitting across chemically distinct compounds. To answer this question, we formulate a unified symmetry framework based on the minimal set of order parameters required to describe the coupled magnetic and orbital degrees of freedom.

\begin{figure}[b!]
	\centering
	\includegraphics[width=0.45\textwidth]{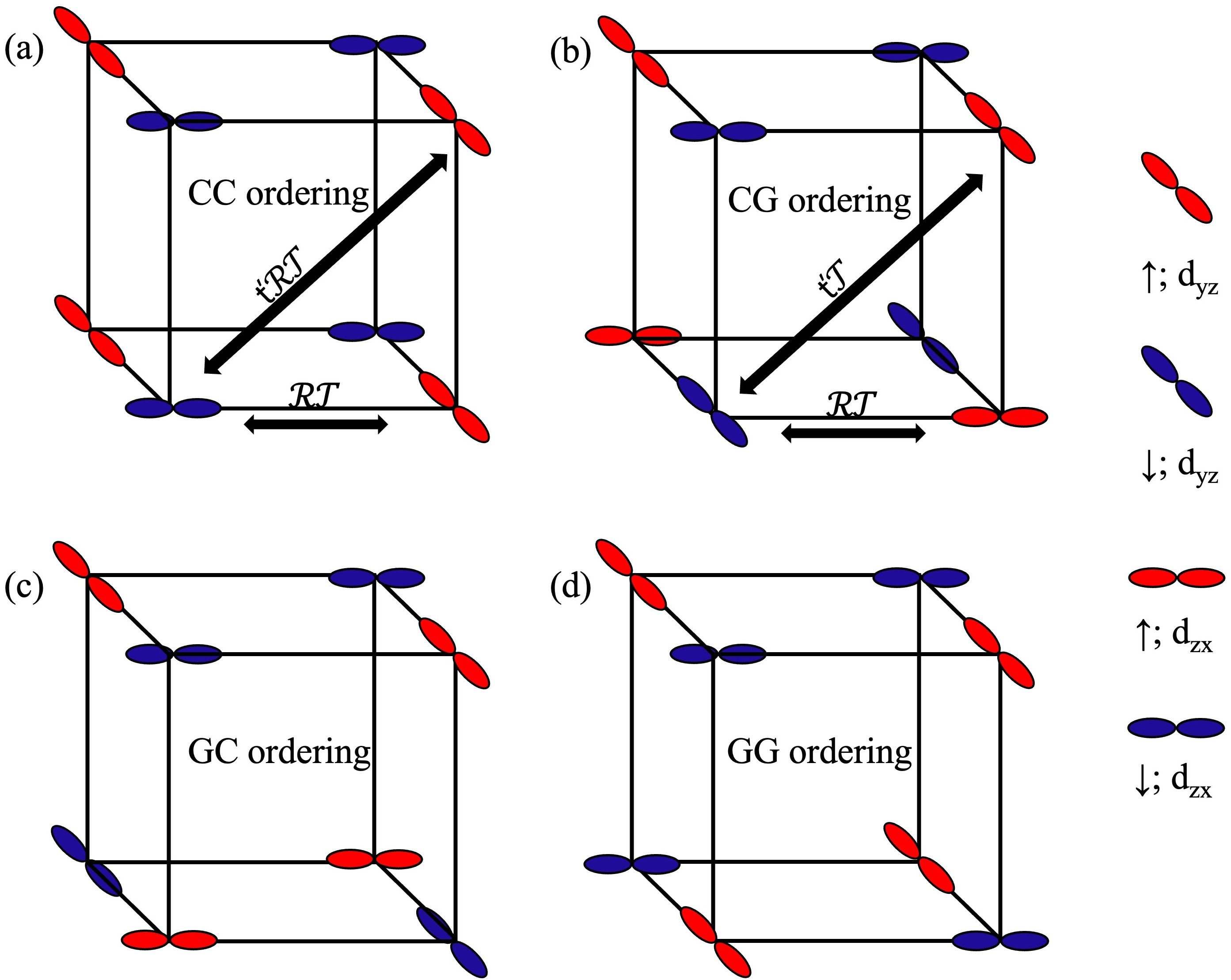}
	\caption{Schematic representation of the possible relative arrangements of orbital ordering (OO) and antiferromagnetic (AFM) order in the bulk structure: (a) CC, (b) CG, (c) GC, and (d) GG configurations. Red and blue lobes denote the occupied $d_{yz}$ and $d_{zx}$ orbital states on opposite magnetic sublattices, highlighting the staggered orbital ordering associated with the antiferromagnetic configuration. Within each layer, the magnetic sublattices are related by the combined rotation--time-reversal symmetry $\mathcal{RT}$. The relative stacking of the AFM and OO order parameters along the crystallographic $c$ direction determines whether the local altermagnetic spin splitting remains uncompensated, giving rise to a bulk altermagnetic state (CC), or is compensated by symmetry-related neighboring layers, resulting in an antialtermagnetic state (CG, GC, and GG).}
	\label{fig:bulk_ordering}
\end{figure}

Remarkably, despite the diversity of crystal structures, electron fillings, and magnetic configurations investigated in this work, all orbital-order-driven altermagnetic states can be described by only two microscopic order parameters: the layer-dependent magnetic order parameter $L_i$ and the staggered orbital order parameter $\Lambda_i$. The former characterizes the antiferromagnetic order within layer $i$, while the latter quantifies the spontaneous staggered orbital polarization arising from the occupation imbalance of the active $d_{xz}/d_{yz}$ orbital manifold. Together, these two order parameters viz. $(L_i,\Lambda_i)$ constitute the minimal microscopic descriptors governing orbital-order-driven altermagnetism.

The magnetic configuration of layer $i$ is described by

\begin{equation}
	L_i=S_A^i-S_B^i,
\end{equation}

where $S_A^i$ and $S_B^i$ denote the spin moments on magnetic sublattices $A$ and $B$, respectively, within the $i^{\mathrm{th}}$ magnetic layer. The corresponding staggered orbital ordering is characterized by

\begin{equation}
	\Lambda_i=\Delta n_A^i-\Delta n_B^i,
\end{equation}

where $\Delta n=n_{xz}-n_{yz}$ represents the orbital polarization arising from the occupation imbalance between the active $d_{xz}$ and $d_{yz}$ orbitals. A finite value of $\Lambda_i$ therefore reflects the spontaneous lifting of the orbital degeneracy, establishing inequivalent orbital occupations on the two magnetic sublattices while preserving the collinear antiferromagnetic order.

As summarized in Fig.~\ref{fig:bulk_ordering}, the possible combinations of the magnetic and orbital order parameters generate four symmetry-allowed stacking configurations. Here, $\mathcal{T}$ denotes the time-reversal operation, $\mathcal{R}$ represents the fourfold spatial rotation relating the magnetic sublattices within an individual layer, and $t^{\prime}$ denotes the translation connecting adjacent magnetic layers along the crystallographic $c$ direction. Independent of the stacking sequence, spontaneous orbital ordering, together with the collinear spin arrangement, establishes the combined rotation--time-reversal symmetry ($\mathcal{RT}$), lifts the translational equivalence of the magnetic sublattices, and generates layer-resolved nonrelativistic spin splitting. The decisive distinction between the four configurations therefore originates not from the local electronic structure but from the symmetry relating neighboring layers.

When both $L_i$ and $\Lambda_i$ exhibit identical stacking (CC configuration) Fig.~\ref{fig:bulk_ordering}(a), adjacent layers remain connected through the combined symmetry $t^{\prime}\mathcal{RT}$, preserving the sign of the layer-resolved spin splitting throughout the crystal. Consequently, the local altermagnetic response survives in the bulk electronic structure, giving rise to a genuine bulk altermagnetic phase.

By contrast, reversing the stacking sequence of either the magnetic or orbital order parameter (CG, GC, and GG configurations) Figs.~\ref{fig:bulk_ordering}(b--c) transforms the interlayer symmetry into $t^{\prime}\mathcal{T}$ or an equivalent symmetry operation. Although each individual layer continues to exhibit finite nonrelativistic spin splitting generated by the local $\mathcal{RT}$ symmetry, the layer-resolved spin splitting reverses sign between neighboring layers, resulting in an exact compensation in the bulk electronic structure. These compensated phases therefore realize antialtermagnetism, in which hidden layer-resolved altermagnetism coexists with globally spin-degenerate bands.

The above analysis establishes a simple and general symmetry criterion for orbital-order-driven altermagnetism. The classification of Type-I ($t^{\prime}\mathcal{T}$) and Type-II ($t^{\prime}\mathcal{T}$) Ruddlesden--Popper stacking configurations is presented in Sec.~I of the SM. While the local $\mathcal{RT}$ symmetry is responsible for generating layer-resolved nonrelativistic spin splitting, the interlayer symmetry uniquely determines its macroscopic manifestation. Structures related by $t^{\prime}\mathcal{RT}$ retain the same spin polarization across neighboring layers and therefore realize bulk altermagnetism, whereas those connected by $t^{\prime}\mathcal{T}$ exhibit an exact compensation of the layer-resolved spin splitting, giving rise to antialtermagnetism. Importantly, this symmetry criterion is independent of the microscopic origin of the orbital ordering and is therefore broadly applicable across transition-metal compounds with diverse electron fillings and crystal structures.

Having established the symmetry conditions governing orbital-order-driven altermagnetism, we next demonstrate that these conditions naturally emerge from a universal crystal-field reconstruction mechanism operating across diverse transition-metal compounds.

\subsection*{Universal crystal-field design principle}
Having established the symmetry conditions required for orbital-order-driven altermagnetism, we now identify the common microscopic mechanism responsible for realizing these conditions across chemically distinct compounds. All investigated compounds exhibit crystal-field reconstruction within the same active $d_{xz}/d_{yz}$ orbital manifold, producing a finite staggered orbital order parameter ($\Lambda_i$), as summarized in Table~\ref{tab:materials}. At first glance, the compounds investigated in this work, appear to be remarkably diverse, encompassing different transition-metal ions, crystal structures, magnetic moments, and electron fillings. Nevertheless, Table~\ref{tab:materials} reveals that they all follow the same sequence of microscopic events, pointing to a common crystal-field design principle.

Structural relaxation first reconstructs the local crystal-field environment, lifting the near-degeneracy of the transition-metal $t_{2g}$ orbitals. The resulting crystal-field splitting selectively stabilizes one orbital while destabilizing another, producing a staggered occupation imbalance between the $d_{xz}$ and $d_{yz}$ orbitals on neighboring magnetic sublattices. This spontaneous orbital polarization generates a finite orbital order parameter $\Lambda_i$, thereby breaking the translational equivalence of the magnetic sublattices while preserving the underlying collinear antiferromagnetic order described by $L_i$.

As summarized in Table~\ref{tab:materials}, this microscopic mechanism is remarkably robust across transition-metal compounds spanning $d^{1}$, $d^{2}$, low-spin $d^{4}$ and $d^{5}$, and high-spin $d^{6}$ and $d^{7}$ electronic configurations. Although these systems possess substantially different crystal structures, magnetic moments, and exchange interactions, they all develop staggered orbital ordering within the same active $d_{xz}/d_{yz}$ orbital manifold. This observation demonstrates that orbital-order-driven altermagnetism is governed primarily by crystal-field reconstruction rather than by a particular electron filling or chemical composition.

An equally important conclusion emerging from Table~\ref{tab:materials} is the direct correspondence between interlayer symmetry and the resulting magnetic phase. Whenever adjacent layers are related by $t^{\prime}\mathcal{RT}$, the layer-resolved nonrelativistic spin splitting survives throughout the crystal, giving rise to a bulk altermagnetic state. Conversely, structures connected by $t^{\prime}\mathcal{T}$ exhibit an exact compensation of the layer-resolved spin splitting and consequently realize antialtermagnetism. Thus, crystal-field reconstruction determines the local electronic instability, whereas the interlayer symmetry dictates its macroscopic manifestation.

\subsection*{Representative realization in a $d^{1}$ perovskite: SrVO$_3$}

\begin{figure}[ht]
	\centering
	\includegraphics[width=\columnwidth]{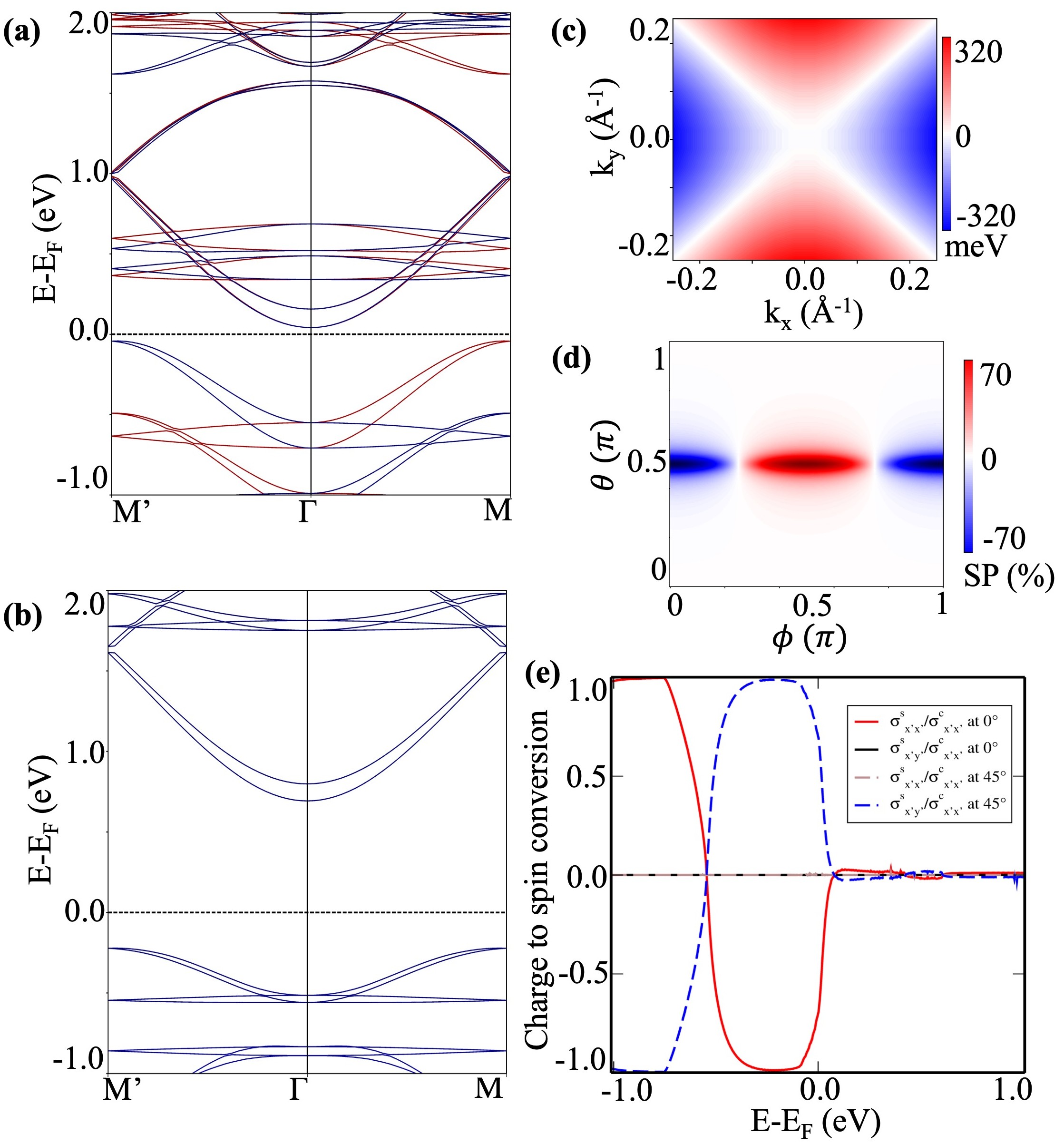}
	\caption{Representative realization of orbital-order-driven altermagnetism in the $d^{1}$ perovskite SrVO$_3$. (a) Spin-resolved electronic band structure of the CC orbital-ordering configuration, exhibiting pronounced bulk nonrelativistic spin splitting (NRSS) along the $M^{\prime}$--$\Gamma$--$M$ path. (b) Spin-resolved electronic band structure of the CG orbital-ordering configuration, where the bulk electronic bands remain spin degenerate owing to symmetry-enforced antialtermagnetic compensation. (c) Momentum-resolved spin-splitting energy, $\Delta E(\mathbf{k})=E_{\uparrow}(\mathbf{k})-E_{\downarrow}(\mathbf{k})$,	around the valence-band maximum (VBM), revealing the characteristic $d$-wave spin-splitting texture. (d) Angular dependence of the spin polarization, $SP_{\hat{n}}$, demonstrating the symmetry-imposed anisotropic spin response. (e) Energy-dependent longitudinal ($\sigma^{s}_{x^{\prime}x^{\prime}}/\sigma^{c}_{x^{\prime}x^{\prime}}$) and transverse ($\sigma^{s}_{x^{\prime}y^{\prime}}/\sigma^{c}_{x^{\prime}x^{\prime}}$) charge-to-spin conversion efficiencies, highlighting efficient spin-current generation near the band edges. Together, these results establish SrVO$_3$ as a representative realization of the proposed crystal-field design principle, linking staggered orbital ordering to nonrelativistic spin splitting, anisotropic spin polarization, and charge-to-spin conversion.}
	\label{fig:SrVO3}
\end{figure}

To demonstrate the universal crystal-field design principle in a realistic material, we first consider the simplest representative $d^{1}$ perovskite, SrVO$_3$. With a single electron occupying the partially filled $t_{2g}$ manifold, SrVO$_3$ provides an ideal platform for illustrating how staggered orbital ordering gives rise to orbital-order-driven altermagnetism. As the simplest representative of the universal design principle established above, it serves as a benchmark for understanding the microscopic origin of orbital-order-driven altermagnetism before extending the analysis to more complex electron fillings.

The calculated spin-resolved electronic band structures for the CC and CG orbital-ordering configurations are presented in Figs.~\ref{fig:SrVO3}(a) and \ref{fig:SrVO3}(b), respectively. In the CC configuration, the intralayer $\mathcal{RT}$ symmetry, together with the interlayer $t^{\prime}\mathcal{RT}$ symmetry, preserves the sign of the nonrelativistic spin splitting throughout the crystal. This gives rise to a bulk altermagnetic state with pronounced spin splitting along the $M^{\prime}$--$\Gamma$--$M$ path. In contrast, the CG configuration modifies the interlayer symmetry to $t^{\prime}\mathcal{T}$, causing the spin splitting generated in neighboring layers to cancel exactly (see SM Fig.~S2(a,b)). Consequently, the bulk electronic bands remain spin degenerate despite the persistence of local orbital ordering, establishing the compensated antialtermagnetic state predicted by our symmetry framework.

The momentum-space signature of the altermagnetic state is illustrated in Fig.~\ref{fig:SrVO3}(c), which shows the spin-splitting energy
\[
\Delta E(\mathbf{k}) =
E_{\uparrow}(\mathbf{k})-
E_{\downarrow}(\mathbf{k}),
\]
around the valence-band maximum. The alternating positive and negative sectors exhibit the characteristic $d$-wave symmetry expected for orbital-order-driven altermagnetism, directly reflecting the symmetry constraints imposed by the orbital ordering, with a maximum spin splitting of 320 meV.

The resulting spin polarization at the Fermi energy is summarized in Fig.~\ref{fig:SrVO3}(d). Its angular dependence exhibits symmetry-protected nodal directions and alternating regions of positive and negative spin polarization, in agreement with the analytical symmetry analysis [see SM section 3].

Finally, the experimentally relevant consequence of the orbital-order-driven altermagnetic state is demonstrated in Fig.~\ref{fig:SrVO3}(e), which presents the longitudinal and transverse charge-to-spin conversion efficiencies as functions of energy. As a $d$-wave altermagnet, SrVO$_3$ exhibits distinct spin responses depending on the direction of the applied electric field. Because the spin splitting vanishes along the symmetry-protected $x+y$ and $x-y$ directions [white lines in Fig.~\ref{fig:SrVO3}(c)], an electric field applied along the $x$ direction generates a longitudinal spin current, whereas fields applied along the $x+y$ or $x-y$ directions produce a transverse spin current. The general expressions for the longitudinal and transverse spin conductivity, together with the corresponding charge-to-spin conversion for an arbitrary electric-field direction, are derived in Sec.~S3 of the Supplemental Material. Large conversion efficiencies emerge near the band edges, demonstrating that the crystal-field-induced nonrelativistic spin splitting directly enables efficient electrical generation of spin currents without requiring spin--orbit coupling. These results establish SrVO$_3$ as a representative realization of the proposed crystal-field design principle, demonstrating how staggered orbital ordering gives rise to nonrelativistic spin splitting, which in turn produces the characteristic $d$-wave altermagnetic symmetry, anisotropic spin polarization, and efficient charge-to-spin conversion.

\subsection*{Extending the design principle beyond the $d^{1}$ configuration}

To demonstrate that the proposed crystal-field design principle is not restricted to the simplest $d^{1}$ electronic configuration, we next consider the representative $d^{4}$ perovskite SrRuO$_3$. Despite its substantially different orbital occupancy, our symmetry framework predicts that staggered orbital ordering should produce bulk nonrelativistic spin splitting when the intralayer $\mathcal{RT}$ and interlayer $t^{\prime}\mathcal{RT}$ symmetries are simultaneously preserved.

\begin{figure}[ht]
	\centering
	\includegraphics[width=\columnwidth]{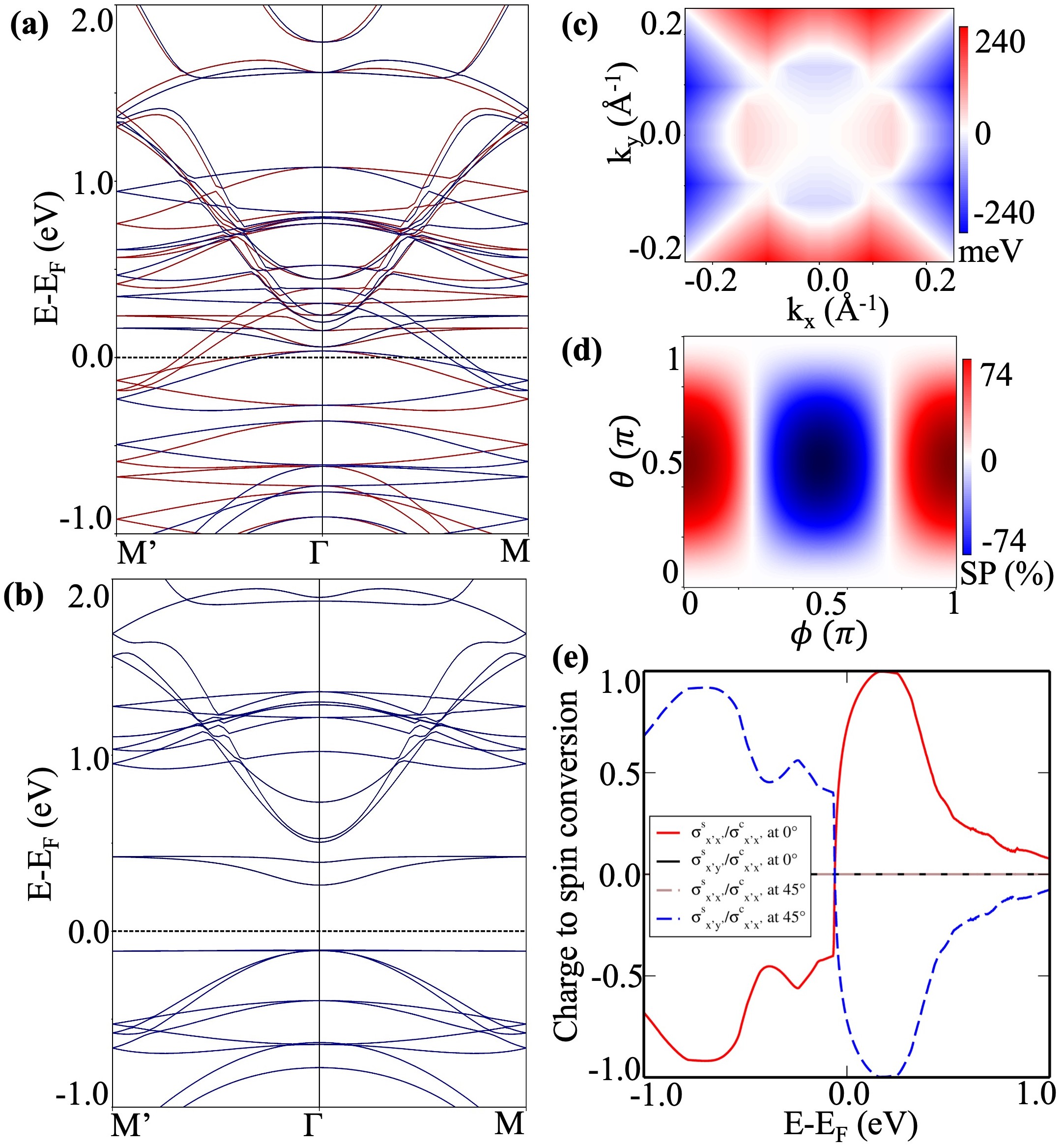}
	\caption{Representative realization of orbital-order-driven altermagnetism in the $d^{4}$ perovskite SrRuO$_3$. (a) Spin-resolved electronic band structure of the CC orbital-ordering configuration, exhibiting pronounced bulk nonrelativistic spin splitting (NRSS) along the $M^{\prime}$--$\Gamma$--$M$ path. (b) Spin-resolved electronic band structure of the CG orbital-ordering configuration, where the bulk electronic bands remain spin degenerate owing to symmetry-enforced antialtermagnetic compensation. (c) Momentum-resolved spin-splitting energy, $\Delta E(\mathbf{k})=E_{\uparrow}(\mathbf{k})-E_{\downarrow}(\mathbf{k})$, around the conduction-band minimum (CBm), revealing the characteristic $d$-wave spin-splitting texture. (d) Angular dependence of the spin polarization, $SP_{\hat{n}}$, demonstrating the symmetry-imposed anisotropic spin response. (e) Energy-dependent longitudinal ($\sigma^{s}_{x^{\prime}x^{\prime}}/\sigma^{c}_{x^{\prime}x^{\prime}}$) and transverse ($\sigma^{s}_{x^{\prime}y^{\prime}}/\sigma^{c}_{x^{\prime}x^{\prime}}$) charge-to-spin conversion efficiencies, highlighting efficient spin-current generation near the band edges. Together, these results demonstrate that the proposed crystal-field design principle extends beyond the $d^{1}$ configuration and remains applicable to transition-metal systems with substantially different electron fillings.}
	\label{fig:SrRuO3}
\end{figure} 

\begin{figure*}[ht]
	\centering
	\includegraphics[width=\linewidth]{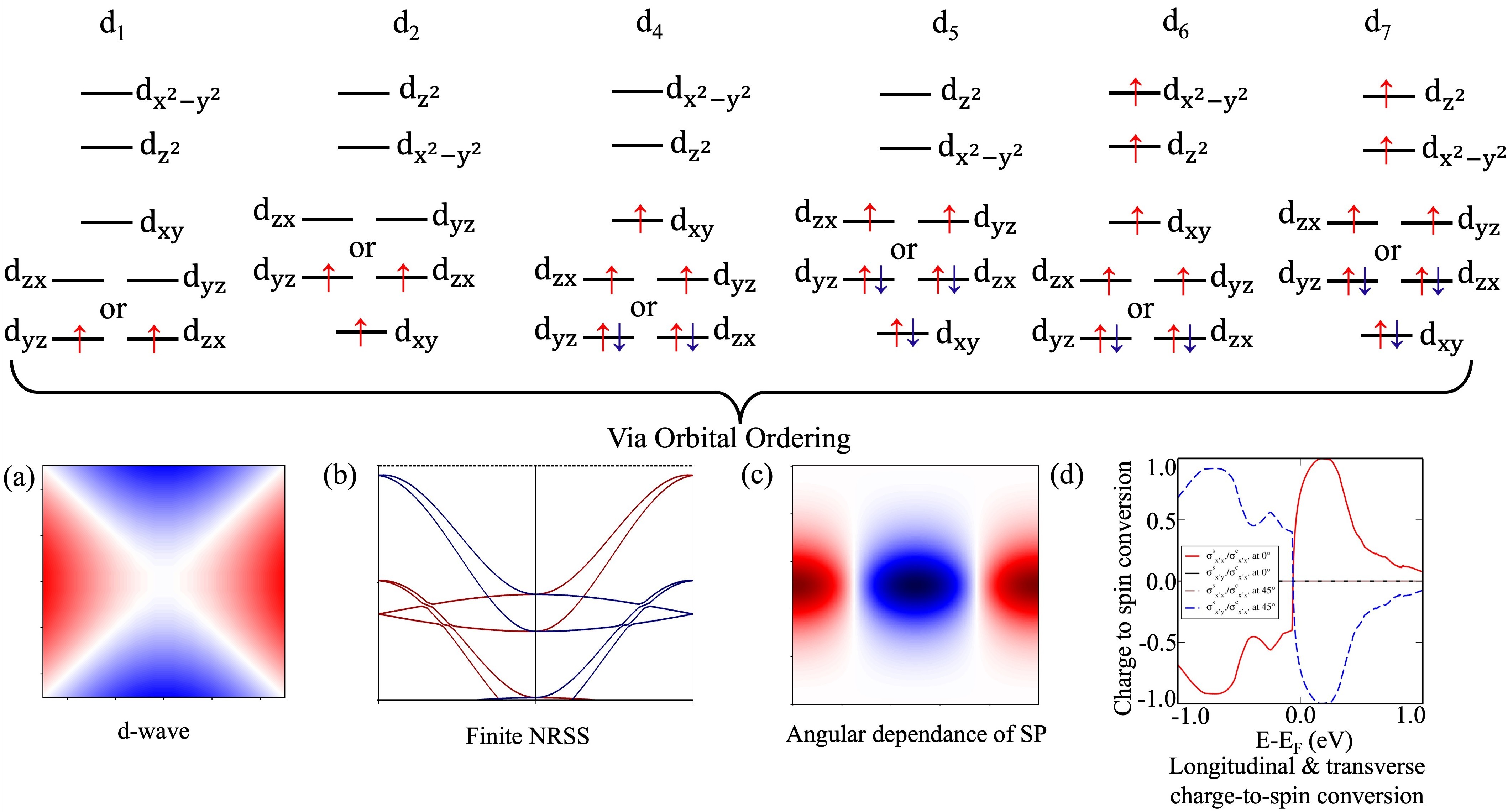}
	\caption{Universal crystal-field design principle for orbital-order-driven altermagnetism.The upper panel summarizes the evolution of crystal-field splitting and orbital occupancy for representative transition-metal electron fillings, illustrating how crystal-field reconstruction generates staggered orbital ordering and the corresponding orbital order parameter $\Lambda_i$. The lower panels summarize the universal physical consequences of this mechanism: (a) emergence of finite nonrelativistic spin splitting (NRSS), (b) characteristic $d$-wave momentum-space spin texture, (c) anisotropic angular spin polarization, and (d) efficient charge-to-spin conversion. Together, these results establish crystal-field engineering as a universal route to realizing orbital-order-driven altermagnetism across transition-metal compounds with diverse electron fillings.}
	\label{fig:Universal}
\end{figure*}

The calculated spin-resolved electronic band structure of the CC orbital-ordering configuration, which preserves the $t^{\prime}\mathcal{RT}$ symmetry, is presented in Fig.~\ref{fig:SrRuO3}(a). Pronounced bulk nonrelativistic spin splitting is observed along the $M^{\prime}$--$\Gamma$--$M$ path, confirming that the microscopic mechanism responsible for orbital-order-driven altermagnetism remains operative despite the substantially different orbital occupancy of the $d^4$ configuration. In contrast, the CG configuration [Fig.~\ref{fig:SrRuO3}(b)] preserves the $t^{\prime}\mathcal{T}$ symmetry, causing the spin splitting generated in neighboring layers to cancel exactly (see SM Fig.~S9(a,b)). Consequently, the bulk electronic bands remain spin degenerate despite the persistence of local orbital ordering, establishing the compensated antialtermagnetic state predicted by our symmetry framework. The corresponding momentum-resolved spin-splitting energy, shown in Fig.~\ref{fig:SrRuO3}(c), exhibits the characteristic alternating positive and negative sectors expected for a $d$-wave altermagnetic state, consistent with the symmetry analysis and reaching a maximum magnitude of 240~meV.

The anisotropic spin response is illustrated in Fig.~\ref{fig:SrRuO3}(d). Similar to the $d^{1}$ case, the spin polarization follows the symmetry-imposed angular dependence, reaching its maximum along the principal directions and vanishing along the symmetry-protected nodal directions. This characteristic behavior demonstrates that the spin response is governed primarily by the orbital-ordering symmetry rather than by the specific electron filling.

Finally, Fig.~\ref{fig:SrRuO3}(e) presents the longitudinal and transverse charge-to-spin conversion efficiencies as functions of energy. The direction-dependent spin response follows the same symmetry-governed behavior established for SrVO$_3$, with the general expressions for an arbitrary electric-field direction derived in Sec.~S3 of the Supplemental Material. Large conversion efficiencies are again obtained near the band edges, indicating efficient electrical generation of spin currents without relying on spin--orbit coupling. Together with the representative $d^{1}$ example discussed above, these results demonstrate that orbital-order-driven altermagnetism constitutes a general consequence of the proposed crystal-field design principle rather than a material-specific phenomenon.
The corresponding results for the remaining electron fillings, bulk compounds, and the complete Ruddlesden--Popper homologous series are presented in the Supplemental Material (Secs.~S4--S9), confirming the universal applicability of the proposed mechanism.

\subsection*{Generality of the crystal-field design principle}

Having established the microscopic mechanism in the representative $d^{1}$ and $d^{4}$ perovskites, we now summarize the universal crystal-field design principle governing orbital-order-driven altermagnetism. Fig.~\ref{fig:Universal} illustrates that, despite the diversity of transition-metal electron fillings and Ruddlesden--Popper stacking sequences, all investigated systems follow the same microscopic pathway. Structural relaxation reconstructs the local crystal-field environment, activating the nearly degenerate $d_{xz}/d_{yz}$ orbital manifold and driving spontaneous staggered orbital ordering characterized by a finite orbital order parameter $\Lambda_i$.

This orbital reconstruction constitutes the common microscopic origin of nonrelativistic spin splitting across all investigated compounds. Independent of the specific electron filling, the resulting altermagnetic state exhibits the same characteristic $d$-wave momentum-space spin-split energy texture together with a strongly anisotropic spin response and efficient charge-to-spin conversion, as schematically summarized in the lower panels of Fig.~\ref{fig:Universal}. These features therefore represent universal experimental fingerprints of orbital-order-driven altermagnetism rather than material-specific properties.

The representative systems discussed in the main text demonstrate the operation of this mechanism in the simplest $d^{1}$ and a higher-filling $d^{4}$ configurations. The corresponding results for the remaining electron fillings, together with the complete Ruddlesden--Popper stacking analysis and additional bulk compounds, are presented in the Supplemental Material (Secs.~S4--S9). Collectively, these results establish crystal-field engineering as a predictive and transferable strategy for discovering and designing orbital-order-driven altermagnets across a broad range of transition-metal compounds.

\section*{Conclusions}
In summary, we have established a universal crystal-field design principle for orbital-order-driven altermagnetism. By introducing the layer-dependent magnetic and orbital order parameters, $L_i$ and $\Lambda_i$, we developed a unified symmetry framework that distinguishes bulk altermagnetic and antialtermagnetic phases according to their interlayer symmetry. This framework demonstrates that while the local $\mathcal{RT}$ symmetry generates layer-resolved nonrelativistic spin splitting, the interlayer symmetry uniquely determines whether this spin splitting survives or is compensated in the bulk electronic structure.

First-principles calculations reveal that spontaneous crystal-field reconstruction consistently activates a common active $d_{xz}/d_{yz}$ orbital manifold, driving staggered orbital ordering across a broad range of transition-metal compounds with diverse electron fillings. The representative $d^{1}$ and $d^{4}$ perovskites demonstrate that this common microscopic mechanism produces the characteristic $d$-wave nonrelativistic spin splitting, anisotropic spin polarization, and efficient charge-to-spin conversion, while the Supplemental Material establishes its generality across additional electron fillings, crystal structures, and dimensionalities.

Our results elevate orbital ordering from a material-specific phenomenon to a predictive design principle for altermagnetism. By linking crystal-field engineering to orbital ordering, symmetry-governed spin splitting, and spin transport, this work provides a transferable framework for discovering and designing altermagnetic materials, opening new opportunities for spintronic technologies operating without reliance on spin--orbit coupling.

\section*{Acknowledgments}  
S.P. acknowledges PMRF, India, for the Research fellowship [Grant No. 1403227]. S. B. acknowledges financial support from SERB under a core research grant (Grant no. CRG/2019/000647) to set up his High Performance Computing (HPC) facility “Veena” at IIT Delhi for computational resources. We acknowledge "Tejas" IoE HPC facility of Dept. of Physics, IIT Delhi. 

\bibliography{manuscript}

@article{vsmejkal2022beyond,
	title = {Beyond Conventional Ferromagnetism and Antiferromagnetism: A Phase with Nonrelativistic Spin and Crystal Rotation Symmetry},
	author = {\ifmmode \check{S}\else \v{S}\fi{}mejkal, Libor and Sinova, Jairo and Jungwirth, Tomas},
	journal = {Phys. Rev. X},
	volume = {12},
	issue = {3},
	pages = {031042},
	numpages = {16},
	year = {2022},
	month = {Sep},
	publisher = {American Physical Society},
	doi = {10.1103/PhysRevX.12.031042},
	url = {https://link.aps.org/doi/10.1103/PhysRevX.12.031042}
}

@article{jungwirth2024altermagnets,
	title={Altermagnets and beyond: Nodal magnetically-ordered phases},
	author={Jungwirth, Tomas and Fernandes, Rafael M and Sinova, Jairo and Smejkal, Libor},
	journal={arXiv preprint arXiv:2409.10034},
	year={2024}
}

@article{gonzalez2021efficient,
	title = {Emerging Research Landscape of Altermagnetism},
	author = {\ifmmode \check{S}\else \v{S}\fi{}mejkal, Libor and Sinova, Jairo and Jungwirth, Tomas},
	journal = {Phys. Rev. X},
	volume = {12},
	issue = {4},
	pages = {040501},
	numpages = {27},
	year = {2022},
	month = {Dec},
	publisher = {American Physical Society},
	doi = {10.1103/PhysRevX.12.040501},
	url = {https://link.aps.org/doi/10.1103/PhysRevX.12.040501}
}

@article{bailing,
	author = {Bai, Ling and Feng, Wanxiang and Liu, Siyuan and Šmejkal, Libor and Mokrousov, Yuriy and Yao, Yugui},
	title = {Altermagnetism: Exploring New Frontiers in Magnetism and Spintronics},
	journal = {Advanced Functional Materials},
	volume = {34},
	number = {49},
	pages = {2409327},
	doi = {https://doi.org/10.1002/adfm.202409327},
	year = {2024}
}

@article{Denisov2024PRB_NRSDynamics,
	author = {Denisov, K. S. and \v{Z}uti\'{c}, I.},
	title = {Anisotropic Spin Dynamics in Antiferromagnets with a Nonrelativistic Spin Splitting},
	journal = {Phys. Rev. B},
	volume = {110},
	pages = {L180403},
	year = {2024},
	doi = {10.1103/PhysRevB.110.L180403}
}

@article{Fang2024PRL_QGAM,
	author = {Fang, Y. and Cano, J. and Ghorashi, S. A. A.},
	title = {Quantum Geometry Induced Nonlinear Transport in Altermagnets},
	journal = {Phys. Rev. Lett.},
	volume = {133},
	pages = {106701},
	year = {2024},
	doi = {10.1103/PhysRevLett.133.106701}
}

@article{w52v-blqm, title = {Anisotropic spin-polarized conductivity in collinear altermagnets}, author = {Dou, Mingbo and Wang, Xianjie and Tao, L. L.}, journal = {Phys. Rev. B}, volume = {111}, issue = {22}, pages = {224423}, numpages = {8}, year = {2025}, month = {Jun}, publisher = {American Physical Society}, doi = {10.1103/w52v-blqm}, url = {https://link.aps.org/doi/10.1103/w52v-blqm} }

@article{sinova2015spin,
	title={Spin Hall effects},
	author={Sinova, Jairo and Valenzuela, Sergio O. and Wunderlich, J. and Back, Christian H. and Jungwirth, Tomas},
	journal={Rev. Mod. Phys.},
	volume={87},
	pages={1213},
	year={2015}
}

@article{vzutic2004spintronics,
  title={Spintronics: Fundamentals and applications},
  author={Žutić, Igor and Fabian, Jaroslav and Das Sarma, S},
  journal={Rev. Mod. Phys.},
  volume={76},
  pages={323},
  year={2004}
}

@article{Pan2024PRL_GST,
	author = {Pan, B. and Zhou, P. and Lyu, P. and Xiao, H. and Yang, X. and Sun, L.},
	title = {General Stacking Theory for Altermagnetism in Bilayer Systems},
	journal = {Phys. Rev. Lett.},
	volume = {133},
	pages = {166701},
	year = {2024},
	doi = {10.1103/PhysRevLett.133.166701}
}

@article{Samanta2025NanoLett_filter,
	author = {Samanta, Kartik and Shao, Ding-Fu and Tsymbal, Evgeny Y.},
	title = {Spin Filtering with Insulating Altermagnets},
	journal = {Nano Letters},
	year = {2025},
	volume = {25},
	number = {8},
	pages = {3150--3156},
	doi = {10.1021/acs.nanolett.4c05672}
}

@article{feng2022anomalous,
	title={An anomalous Hall effect in altermagnetic ruthenium dioxide},
	author={Feng, Zexin and Zhou, Xiaorong and {\v{S}}mejkal, Libor and Wu, Lei and Zhu, Zengwei and Guo, Huixin and Gonz{\'a}lez-Hern{\'a}ndez, Rafael and Wang, Xiaoning and Yan, Han and Qin, Peixin and others},
	journal={Nature Electronics},
	volume={5},
	number={11},
	pages={735--743},
	year={2022},
	publisher={Nature Publishing Group UK London}
}

@article{ss_prb,
	title = {Tuning spin currents in collinear antiferromagnets and altermagnets},
	author = {Sheoran, Sajjan and Dev, Pratibha},
	journal = {Phys. Rev. B},
	volume = {113},
	issue = {17},
	pages = {174426},
	numpages = {10},
	year = {2026},
	month = {May},
	publisher = {American Physical Society},
	doi = {10.1103/159t-fq4k},
	url = {https://link.aps.org/doi/10.1103/159t-fq4k}
}

@article{quintin,
	title = {Net and Compensated Altermagnetism from Staggered Orbital Order: Layer-Dependent Spin Splitting in ${\mathrm{Sr}}_{n+1}{\mathrm{Cr}}_{n}{\mathrm{O}}_{3n+1}$},
	author = {Meier, Quintin N. and Carta, Alberto and Ederer, Claude and Cano, Andr\'es},
	journal = {Phys. Rev. Lett.},
	volume = {136},
	issue = {11},
	pages = {116705},
	numpages = {8},
	year = {2026},
	month = {Mar},
	publisher = {American Physical Society},
	doi = {10.1103/mm8t-82q4},
	url = {https://link.aps.org/doi/10.1103/mm8t-82q4}
}

@article{PhysRevLett.132.236701,
	title = {Spontaneous Formation of Altermagnetism from Orbital Ordering},
	author = {Leeb, Valentin and Mook, Alexander and \ifmmode \check{S}\else \v{S}\fi{}mejkal, Libor and Knolle, Johannes},
	journal = {Phys. Rev. Lett.},
	volume = {132},
	issue = {23},
	pages = {236701},
	numpages = {7},
	year = {2024},
	month = {Jun},
	publisher = {American Physical Society},
	doi = {10.1103/PhysRevLett.132.236701},
	url = {https://link.aps.org/doi/10.1103/PhysRevLett.132.236701}
}

@article{pathak2026strain,
	title={Strain-and Field-Tunable Nonrelativistic Spin Splitting and Wave-Symmetry-Dependent Spin Transport in Twisted Bilayer Altermagnets},
	author={Pathak, Shantanu and Bhattacharya, Saswata},
	journal={arXiv preprint arXiv:2602.19713},
	year={2026}
}

@book{callaway2013quantum,
	title={Quantum Theory of the Solid State},
	author={Callaway, Joseph},
	year={2013},
	publisher={Academic Press}
}

@article{Zhang2026NanoLett_hidden,
	author = {Zhang, Tongxie and Yuan, Linding and Rondinelli, James M. and Fertig, H. A. and Zhang, Shixiong},
	title = {Tunable Hidden Altermagnetic Spin Splitting in Layered Ruddlesden--Popper Oxides},
	journal = {Nano Letters},
	year = {2026},
	doi = {10.1021/acs.nanolett.6c00013}
}

@article{kresse1996vasp1,
	title={Efficient iterative schemes for ab initio total-energy calculations using a plane-wave basis set},
	author={Kresse, G. and Furthmüller, J.},
	journal={Physical Review B},
	volume={54},
	pages={11169--11186},
	year={1996}
}

@article{kresse1996vasp2,
	title={Efficiency of ab-initio total energy calculations for metals and semiconductors using a plane-wave basis set},
	author={Kresse, G. and Furthmüller, J.},
	journal={Computational Materials Science},
	volume={6},
	pages={15--50},
	year={1996}
}

@article{blochl1994projector,
	title={Projector augmented-wave method},
	author={Bl{\"o}chl, P. E.},
	journal={Physical Review B},
	volume={50},
	pages={17953--17979},
	year={1994}
}

@article{perdew1996generalized,
	title={Generalized gradient approximation made simple},
	author={Perdew, J. P. and Burke, K. and Ernzerhof, M.},
	journal={Physical Review Letters},
	volume={77},
	pages={3865--3868},
	year={1996}
}

@article{dudarev1998electron,
	title={Electron-energy-loss spectra and the structural stability of nickel oxide: An LSDA+ U study},
	author={Dudarev, Sergei L and Botton, Gianluigi A and Savrasov, Sergey Y and Humphreys, CJ and Sutton, Adrian P},
	journal={Physical Review B},
	volume={57},
	number={3},
	pages={1505},
	year={1998},
	publisher={APS}
}

@article{mostofi2008wannier90,
	title={wannier90: A tool for obtaining maximally-localised Wannier functions},
	author={Mostofi, Arash A and Yates, Jonathan R and Lee, Young-Su and Souza, Ivo and Vanderbilt, David and Marzari, Nicola},
	journal={Computer physics communications},
	volume={178},
	number={9},
	pages={685--699},
	year={2008},
	publisher={Elsevier}
}

@article{tsirkin2021high,
	title={High performance Wannier interpolation of Berry curvature and related quantities with WannierBerri code},
	author={Tsirkin, Stepan S},
	journal={npj Computational Materials},
	volume={7},
	number={1},
	pages={33},
	year={2021},
	publisher={Nature Publishing Group UK London}
}

@article{stokes2005findsym,
	title={Findsym: program for identifying the space-group symmetry of a crystal},
	author={Stokes, H. T. and Hatch, D. M.},
	journal={Journal of Applied Crystallography},
	volume={38},
	pages={237--238},
	year={2005}
}

@article{chen2024_spinsg_prx,
	title={Enumeration and representation theory of spin space groups},
	author={Chen, X. and Ren, J. and Zhu, Y. and Yu, Y. and Zhang, A. and Liu, P. and Li, J. and Liu, Y. and Li, C. and Liu, Q.},
	journal={Phys. Rev. X},
	volume={14},
	pages={031038},
	year={2024}
}

@article{chen2024_spingroup_prb110,
	title={Unconventional magnons in collinear magnets dictated by spin space groups},
	author={Chen, Xiaobing and Liu, Yuntian and Liu, Pengfei and Yu, Yutong and Ren, Jun and Li, Jiayu and Zhang, Ao and Liu, Qihang},
	journal={Nature},
	volume={640},
	number={8058},
	pages={349--354},
	year={2025},
	publisher={Nature Publishing Group UK London}
}

@article{xiao2025tensorsymmetry,
	title={TensorSymmetry: a package to get symmetry-adapted tensors disentangling spin-orbit coupling effect and establishing analytical relationship with magnetic order},
	author={Xiao, Rui-Chun and Jin, Yuanjun and Zhang, Zhi-Fan and Feng, Zi-Hao and Shao, Ding-Fu and Tian, Mingliang},
	journal={Computer Physics Communications},
	pages={109872},
	year={2025},
	publisher={Elsevier}
}

@misc{SM,
	note = {See Supplemental Material for computational details and additional results.}
}

@article{aroyo2006bilbao,
	title={Bilbao Crystallographic Server: I. Databases and crystallographic computing programs},
	author={Aroyo, M. I. and others},
	journal={Zeitschrift für Kristallographie},
	volume={221},
	number={1},
	pages={15--27},
	year={2006}
}

@article{aroyo2011crystallography,
	title={Crystallography online: Bilbao Crystallographic Server},
	author={Aroyo, M. I. and others},
	journal={Bulg. Chem. Commun.},
	volume={43},
	pages={183--197},
	year={2011}
}
\end{document}